\documentclass[
    ,final            
  ]
  {aipproc}

\layoutstyle{6x9}

\usepackage{amsfonts}
\usepackage{amsmath}
\usepackage{amssymb}
\usepackage{graphicx}
\usepackage{color}

\def \L{L}
\def \R{R}
\def\||{{\, || \,}}

\def\bor{\mathop{\mathord{\lor}\!\!\!\raise4pt\hbox{$\scriptscriptstyle 2$}\,}}
\def\band{\mathop{\mathord{\land}\!\!\!\lower2pt\hbox{$\scriptscriptstyle 2$}\,}}

\begin{document}

\title{The Problem of Motion: \\The Statistical Mechanics of Zitterbewegung}


\author{Kevin H. Knuth}{
  address={Departments of Physics and Informatics\\
        University at Albany (SUNY), Albany NY, USA\\
        email: kknuth@albany.edu \\ web: http://knuthlab.rit.albany.edu}
}

\begin{abstract}
Around 1930, both Gregory Breit and Erwin Schr\"{o}dinger showed that the eigenvalues of the velocity of a particle described by wavepacket solutions to the Dirac equation are simply $\pm$c, the speed of light.  This led Schr\"{o}dinger to coin the term \emph{Zitterbewegung}, which is German for ``trembling motion'', where all particles of matter (fermions) zig-zag back-and-forth at only the speed of light.  The result is that any finite speed less than $c$, including the state of rest, only makes sense as a long-term average that can be thought of as a drift velocity.  In this paper, we seriously consider this idea that the observed velocities of particles are time-averages of motion at the speed of light and demonstrate how the relativistic velocity addition rule in one spatial dimension is readily derived by considering the probabilities that a particle is observed to move either to the left or to the right at the speed of light.
\end{abstract}

\maketitle


\section{Introduction}
Over two thousand years ago the doctrine of Parmenides, which held that motion is an illusion, resulted in a series of debates that led Zeno of Elea to produce at least nine paradoxes related to the nature of motion.  While several of these paradoxes have been continuously debated throughout the intervening centuries, the work of Galileo and Newton resulted in the fields of kinematics and dynamics, respectively, which for the first time in human history provided predictive power regarding the motion of objects in a wide variety of situations.  As a result, the attentions of the natural philosophers (physicists) were directed toward better describing, predicting and controlling such motions and away from apparently less productive philosophical matters.  As physics advanced, Newton's concepts of absolute space and time gave way to Einstein's relativity where space and time became intertwined and motion was found to be limited by an absolute speed limit---the speed of light, which is constant for all observers.

The advent of relativistic quantum mechanics saw the introduction of the Dirac equation \cite{Dirac:1928}, which describes the behavior of a single particle of matter called a fermion.  In addition to describing particles, such as an electron, the Dirac equation made a startling prediction---the existence of antimatter.  This success resulted in a dramatic change in how we conceive of matter and energy.  However, it was found that the Dirac equation made another interesting prediction.  Both Breit \cite{Breit:1928} and Schr\"{o}dinger \cite{Schrodinger1930:kraftefreie} found that the velocity eigenvalues of a particle described by a wavepacket can only be plus or minus the speed of light: $\pm c$.  That is, at the quantum level, a particle can \textbf{only} be observed to go at the speed of light.  This is challenging in the sense that since the acceptance of relativity, there is no theory as to how to deal with particles moving at the speed of light, since relativity says that massive particles must move at speeds less than the speed of light.  In contrast, the Dirac theory requires that the finite speed of a particle, $v < c$, including rest $v = 0$, be described as the \emph{average} speed of a particle that is zig-zagging back-and-forth at the speed of light.  Thus any observed macroscopic speed is a resultant drift velocity.  Schr\"{o}dinger called this zig-zagging motion, \emph{Zitterbewegung}, which is German for trembling-motion. We refer to this phenomenon simply as \emph{zitter}.

The angular frequency at which a particle zitters is given by twice the de Broglie electron clock rate as determined in his 1924 dissertation \cite{Hestenes:2008electron}
\begin{equation}
\omega = \frac{2 mc^2}{\hbar} \approx 10^{21} \mbox{Hz},
\end{equation}
which corresponds to an immeasurably small time.  This corresponds to a characteristic length of
\begin{equation}
\lambda = c/{\omega} = \frac{\hbar}{2mc} \approx 10^{-13} \mbox{m},
\end{equation}
which is half the reduced Compton wavelength---the finest resolution with which the position of an electron can be observed.
For this reason, zitter has never been directly observed. As a result, the phenomenon of Zitterbewegung is barely mentioned even in the most complete of physics texts. However, there is some experimental evidence that it may be a real effect based on an electron channeling experiment in silicon that detected a resonance at the electron clock rate \cite{Gouanere+etal:2005:channeling}\cite{Catillon+etal:2008clock}, which was explained theoretically by Hestenes \cite{Hestenes:2010zitterbewegung}.  In addition, zitter-like behavior has been observed in trapped ions \cite{Gerritsma+etal:2010quantum} and atoms in a Bose-Einstein condensate \cite{Leblanc+etal:2013:zitter}\cite{Qu+etal:2013:zitter}, which are known to obey the Dirac equation.

If zitter is a real effect, then it could potentially significantly impact how we think about motion.  For example, if particles zitter, then at the quantum level there is no such thing as rest.  A concept of rest would be reasonable only in the sense of long-term averages.  What this does to the concept of a rest frame, which is fundamental to spacetime physics, as we know it, is unclear.  Furthermore, it is important to note that zitter would describe the kinematics of the particle, but not the dynamics, since it is not well-understood what would cause the particle to undergo reversals in direction.\footnote{Keep in mind that it is not known if particles actually zitter.  For this reason, the cause of reversals is subject to speculation. There have been suggestions that it is related to the following: spin, scattering off the Higgs field (mass), as well as direct particle-particle influences \cite{Knuth:Info-Based:2014}.}

In this paper, I consider what happens if one takes zitter seriously.  With a single assumption
I will demonstrate that the concept of zitter in 1+1 dimensions is consistent with the concept of motion in special relativity on average.  The result is a statistical concept of motion where probability theory plays a central role.

\section{Velocity in 1+1 Dimensions}
In 1+1 dimensions, a particle can either move to the left at speed $c$ or to the right at speed $c$, that is, with a velocity of $\pm c$.  In this treatment, we will choose natural units where the speed of light $|c| = 1$.  The Zitterbewegung effect will be treated statistically where in 1+1 dimensions we need only consider the probability that the particle moves to the right, $Pr(\R)$, and the probability that it moves to the left, $Pr(\L)$.  We then have the normalization condition
\begin{equation}
Pr(\R) + Pr(\L) = 1
\end{equation}
as well as the expected value of the speed, denoted by $v$, which is given by
\begin{align}
v &= (+1)\times Pr(\R) + (-1) \times Pr(\L)\\
&= Pr(\R) - Pr(\L).
\end{align}
The sum and difference of these two expressions enables one to write the probabilities in terms of the expected value of the speed
\begin{eqnarray}
Pr(\R) &=& \frac{1}{2}(1+v) \label{eq:Pr(R)} \\
Pr(\L) &=& \frac{1}{2}(1-v). \label{eq:Pr(L)}
\end{eqnarray}
Note that when the particle is at rest on average, $v = 0$, we have that the particle has an equal probability of moving left or right so that $Pr(\R) = Pr(\L) = 1/2$.  At the other extreme, when $v = \pm 1$, we have that the particle never moves to the left/right so that either $Pr(\R) = 0$ and $Pr(L) = 1$ or $Pr(\R) = 1$ and $Pr(L) = 0$.

\section{Zittering Observers}
We now consider that a particle $\Pi$ is seen by observers that are themselves zittering.  Say that observer $O$ perceives the particle $\Pi$ to have an average velocity $v$.  Writing the probability that $O$ observes $\Pi$ going to the right as $Pr(\Pi = \R | O)$ and the probability that $O$ observes $\Pi$ going to the left as $Pr(\Pi = \L | O)$, we have that
\begin{eqnarray}
Pr(\Pi = \R | O) &=& \frac{1}{2}(1+v) \label{eq:first}\\
Pr(\Pi = \L | O) &=& \frac{1}{2}(1-v).
\end{eqnarray}

We now consider another observer $O'$ that has an average non-zero motion, $u$, with respect to the original observer $O$ so that $O'$ has unequal probabilities of being observed to move to the left or to the right by observer $O$:
\begin{eqnarray}
Pr(O' = \R | O) &=& \frac{1}{2}(1+u) \\
Pr(O' = \L | O) &=& \frac{1}{2}(1-u).\label{eq:last}
\end{eqnarray}

\begin{figure}
\centering
\includegraphics[scale = 0.22]{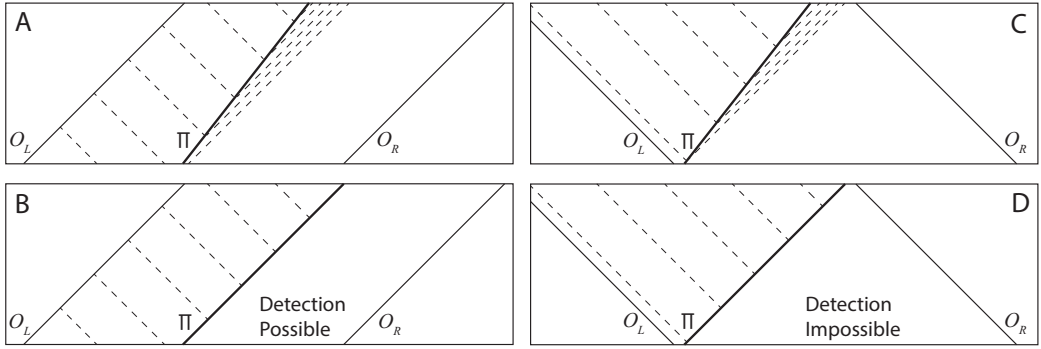}
\caption{This figure provides space-time diagrams illustrating the development of the second assumption that an observer moving at the speed of light can only detect another object moving at the speed of light if they are moving in the same direction (as viewed by a third observer). A. Illustrates an object $\Pi$ (thick solid line) moving at high speed to the right that emits light pulses to the front and back (dashed lines) that are observed by two observers, $O_L$ and $O_R$, (solid lines) moving to the right at the speed of light.  Note that only the observer $O_L$ behind the object can detect the light pulses.  B. This figure is similar to A above, except that the object $\Pi$ is now moving \emph{at} the speed of light.  Again only the observer moving in the same direction behind the particle can detect the light pulses.  In this case, detection is possible.  C. This illustrates observers moving in a direction opposite to the object $\Pi$, which moves at a high speed less than the speed of light. In this case $O_R$ detects the light pulses piling up, whereas observer $O_L$ cannot detect the light pulses.  D. This figure is similar to C except that the object $\Pi$ is now moving at the speed of light.  Note that $O_R$ cannot detect the light pulses while the object moves to the right as they have piled up (unless there is a collision). So that in this case, where the observer moves opposite to the object, it is impossible for the object to be detected.}
\label{fig:zitter}
\end{figure}

Now let us compute the probabilities that the primed observer $O'$ sees the particle moving to the left and right in terms of the probabilities that the unprimed observer $O$ sees the particle as moving to the left or right.  We make two assumptions.  First, we assume that $O'$ cannot observe $\Pi$ to move in a direction opposite to that detected by $O$.  This implies that
\begin{eqnarray}
Pr(\Pi = \R | O') &\propto& Pr(\Pi = \R | O) \\
Pr(\Pi = \L | O') &\propto& Pr(\Pi = \L | O).
\end{eqnarray}

Second, we assume that if $O'$ observes $\Pi$ to be moving to the right/left at the speed of light then $O$ must observe $O'$ to be moving to the right/left at the speed of light (Figure \ref{fig:zitter}).  Now this may sound like a strange assumption, but remember that we have no theory describing how matter behaves when moving \emph{at} the speed of light.  As illustrated in the figure we can imagine that $\Pi$ emits light pulses and is observed by two observers to the right and the left of the particle.  In the limit as the particle velocity approaches the speed of light, the pulses emitted in the forward direction are seen to travel with the particle from the perspective of an outside observer making them undetectable.  However, the light pulses emitted behind the particle can be detected only by an observer behind the particle moving at the speed of light in the same direction as the particle.  It should be noted that while this is an assumption in the present work that requires observers both to the left and right of the particle, we do have a theory of direct particle-particle influence where one can derive this assumption from the projections of event intervals onto observers \cite{Knuth:FQXI2013}\cite{Knuth:Info-Based:2014}.
This assumption implies that
\begin{eqnarray}
Pr(\Pi = \R | O') &\propto& Pr(O' = \R |O)\\
Pr(\Pi = \L | O') &\propto& Pr(O' = \L |O).
\end{eqnarray}

Together these two assumptions lead to the following un-normalized probabilities:
\begin{eqnarray}
Pr(\Pi = \R | O') &\propto& Pr(\Pi = \R | O) Pr(O' = \R |O)\\
Pr(\Pi = \L | O') &\propto& Pr(\Pi = \L | O) Pr(O' = \L |O),
\end{eqnarray}
which when normalized by requiring $Pr(\Pi = \R | O') + Pr(\Pi = \L | O') = 1$ are
\begin{eqnarray}
Pr(\Pi = \R | O') &=& \frac{Pr(\Pi = \R | O) Pr(O' = \R |O)}{Pr(\Pi = \R | O) Pr(O' = \R |O)+Pr(\Pi = \L | O) Pr(O' = \L |O)}\\
Pr(\Pi = \L | O') &=& \frac{Pr(\Pi = \L | O) Pr(O' = \L |O)}{Pr(\Pi = \R | O) Pr(O' = \R |O)+Pr(\Pi = \L | O) Pr(O' = \L |O)}.
\end{eqnarray}

We then have that the speed $w$ of the particle observed by $O'$ is given by
\begin{align}
w &= Pr(\Pi = \R | O') - Pr(\Pi = \L | O')\\
&= \frac{Pr(\Pi = \R | O) Pr(O' = \R |O) - Pr(\Pi = \L | O) Pr(O' = \L | O)}{Pr(\Pi = \R | O) Pr(O' = \R | O) + Pr(\Pi = \L | O) Pr(O' = \L | O)},
\end{align}
which has the expected antisymmetric-over-symmetric form \cite{Knuth+Bahreyni:JMP2014}\cite{Knuth:FQXI2013}\cite{Knuth:Info-Based:2014}.

Using equations (\ref{eq:first})-(\ref{eq:last}), we can rewrite the speed $w$ in terms of the speeds $u$ and $v$
\begin{equation}
w = \frac{\frac{1}{2}(1+v) \frac{1}{2}(1+u) - \frac{1}{2}(1-v) \frac{1}{2}(1-u)}{\frac{1}{2}(1+v) \frac{1}{2}(1+u) + \frac{1}{2}(1-v) \frac{1}{2}(1-u)},
\end{equation}
which when simplified gives the familiar relativistic velocity addition rule
\begin{equation}
w = \frac{u+v}{1+uv}.
\end{equation}

\section{Entropy}

Given that we can describe a particle's motion in 1+1 dimensions with a probability distribution, it is tempting to consider the entropy of that distribution and ask what relationship it may have to the behavior of the particle.

The Shannon entropy for the particle is simply
\begin{equation}
S = - Pr(\R) \log Pr(\R) - Pr(\L) \log Pr(\L)
\end{equation}
where the base of the logarithm only serves to change the units.
It is necessarily observer-dependent, which can be made more explicit by using (\ref{eq:Pr(R)}) and (\ref{eq:Pr(L)}) to rewrite it in terms of the average velocity $\beta$ of the particle
\begin{equation}
S = - \frac{1}{2}(1+\beta) \log \frac{1}{2}(1+\beta) - \frac{1}{2}(1-\beta) \log \frac{1}{2}(1-\beta).
\end{equation}
One can see that the entropy is a maximum, with a value of $S = \log 2$ for a particle at rest $\beta = 0$, and a minimum, with a value of $S = 0$, for a particle moving at the speed of light.  That is, the rest state is the maximum entropy state, since at any instant it is equally probable that the particle is moving to the left or the right.  At the other extreme, a particle moving at the speed of light represents a state of complete certainty.

This suggests that one might consider motion as a statistical mechanical phenomenon.  Consider the particle at rest, which is in a maximum entropy state.  To reduce the entropy of the particle, one must perform work on it.  Doing work reduces the entropy of the particle, and as a result, the particle moves!

It is interesting to note that with some algebra, one can rewrite this entropy completely in terms of special relativistic quantities
\begin{equation}
S = \log{2\gamma} - \beta \log({1+z})
\end{equation}
where $\gamma$ is the relativistic Lorentz factor $\gamma = (1-\beta^2)^{-1/2}$ and $1+z$ is related to the redshift $z$ given by $1+z = \sqrt{\frac{1+\beta}{1-\beta}}$ for motion in the radial direction.

\section{Conclusion}

We have taken the phenomenon of particle zitter seriously, where it is assumed that particles zig-zag back-and-forth at only the speed of light, and have shown that, in 1+1 dimensions, macroscopic motion is consistent with special relativity in terms of the velocity addition rule.  Perhaps it is more interesting that we found that the results can be derived by simply applying probability theory.  This suggests that one can define an entropy for the particle that is necessarily observer-based and perhaps consider motion as a statistical mechanical phenomenon.

Our recent theory of direct particle-particle influence can be shown to reproduce these results directly.  In fact, another paper in this volume by this author and James Lyons Walsh \cite{Walsh+Knuth:acceleration} applies these concepts to demonstrate that influences can act like forces consistent with special relativity.


\begin{theacknowledgments}
This work was funded in part by a grant from the John Templeton Foundation.  I would like to thank Newshaw Bahreyni and James Lyons Walsh as well as Romke Bontekoe, Anthony Garrett and John Skilling for interesting discussions and helpful comments and questions.  I would also like to thank David Hestenes for his comments on our related work and for putting and keeping zitter on our radar.
\end{theacknowledgments}



\bibliographystyle{aipproc}   

\bibliography{C:/Users/KK952431/kevin/files/papers/bibliography/knuth}

\end{document}